\documentclass[12pt]{article}

\usepackage{amsmath}
\usepackage{amssymb}
\usepackage{epsfig}
\usepackage{graphicx}
\usepackage{cite}

\newcommand{\capdef}{}
\newcommand{\mycaption}[2][\capdef]{\renewcommand{\capdef}{#2}%
        \caption[#1]{{\itshape #2}}}
\makeatletter
\renewcommand{\fnum@table}{\textbf{\tablename~\thetable}}
\renewcommand{\fnum@figure}{\textbf{\figurename~\thefigure}}
\makeatother

\newcounter{myenumi}

\renewcommand{\themyenumi}{\roman{myenumi}}
{\end{list}}

\newlength{\myem}
\newcounter{mysubequation}[equation]

\makeatletter
\renewcommand{\section}{\@startsection{section}{1}{0em}{-\baselineskip}%
{\baselineskip}{\normalfont\large\bfseries}}
\renewcommand{\subsection}%
{\@startsection{subsection}{2}{0em}{-0.7\baselineskip}%
{0.7\baselineskip}{\normalfont\bfseries}}
\makeatother

\newcommand{\bi}{\begin{itemize}}
\newcommand{\ei}{\end{itemize}}

\newcommand{\be}{\begin{equation}}
\newcommand{\ee}{\end{equation}}
\newcommand{\bea}{\begin{eqnarray}}
\newcommand{\eea}{\end{eqnarray}}

\newcommand{\ldm}{\Delta m_{31}^2}
\newcommand{\sdm}{\Delta m_{21}^2}
\newcommand{\deltacp}{\delta_{\mathrm{CP}}}
\newcommand{\stheta}{\sin^2 2 \theta_{13}}

\newcommand{\ie}{{\it i.e.}}

\newcommand{\cf}{{\it cf.}}

\newcommand{\eq}{Eq.}

\newcommand{\fig}{Fig.}

\newcommand{\Ref}{Ref.}
\newcommand{\Refs}{Refs.}
\newcommand{\Sec}{Sec.}

\newcommand{\JHFHK}{{\sc JHF-HK}}

\newcommand{\NuFactII}{{\sc NuFact-II}}

\newcommand{\equ}[1]{\eq~(\ref{equ:#1})}
\newcommand{\figu}[1]{\fig~\ref{fig:#1}}

\begin{document}

\begin{titlepage}

\renewcommand{\thefootnote}{\alph{footnote}}

\vspace*{-3.cm}
\begin{flushright}
TUM-HEP-520/03\\
\end{flushright}

\vspace*{0.5cm}

\renewcommand{\thefootnote}{\fnsymbol{footnote}}
\setcounter{footnote}{-1}

{\begin{center}
{\large\bf The role of matter density uncertainties in the analysis of
future neutrino factory experiments} \end{center}}
\renewcommand{\thefootnote}{\alph{footnote}}

\vspace*{.8cm}
\vspace*{.3cm}
{\begin{center} {\large{\sc
                Tommy~Ohlsson\footnote[1]{\makebox[1.cm]{Email:}
                tommy@theophys.kth.se} and
                Walter~Winter\footnote[2]{\makebox[1.cm]{Email:}
                wwinter@ph.tum.de}~
                }}
\end{center}}
\vspace*{0cm}
{\it
\begin{center}

\footnotemark[1]%
Division of Mathematical Physics, Department of
Physics, Royal Institute of Technology (KTH) -- Stockholm Center for
Physics, Astronomy, and Biotechnology (SCFAB), Roslagstullsbacken
11, 106~91~~Stockholm, Sweden 

\footnotemark[2]%
       Institut f\"ur theoretische Physik, Physik--Department,
       Technische Universit\"at M\"unchen (TUM),
       James--Franck--Strasse, 85748~Garching bei M{\"u}nchen, Germany

\end{center}}

\vspace*{1.5cm}

{\Large \bf
\begin{center} Abstract \end{center}  }
Matter density uncertainties can affect the measurements of the
neutrino oscillation parameters at future neutrino factory
experiments, such as the measurements of the mixing parameters
$\theta_{13}$ and $\deltacp$. We compare different matter density
uncertainty models and discuss the possibility to include the matter
density uncertainties in a complete statistical analysis. Furthermore,
we systematically study in which measurements and where in the parameter
space matter density uncertainties are most relevant. We illustrate
this discussion with examples that show the effects as functions of
different magnitudes of the matter density uncertainties. We find 
that matter density uncertainties are especially relevant
for large $\stheta \gtrsim 10^{-3}$. Within the
KamLAND-allowed range, they are most relevant for the precision
measurements of $\stheta$ and $\deltacp$, but less relevant for ``binary''
measurements, such as for the sign of $\ldm$, the sensitivity to
$\stheta$, or the sensitivity to maximal CP violation. In addition,
we demonstrate that knowing the matter density along a specific
baseline better than to about $1\%$ precision means that all measurements
will become almost independent of the matter density uncertainties.
 
\vspace*{.5cm}

\end{titlepage}

\newpage

\renewcommand{\thefootnote}{\arabic{footnote}}
\setcounter{footnote}{0}


\section{Introduction}

Neutrino oscillation physics has entered the era of precision
measurements of the leading atmospheric ($\Delta m_{31}^2$ and
$\theta_{23}$) and solar ($\Delta m_{21}^2$ and $\theta_{12}$)
neutrino oscillation parameters. Especially, the recent results of the
Super-Kamiokande
\cite{Fukuda:1998mi,Fukuda:1998ah},
SNO
\cite{Ahmad:2001an,Ahmad:2002jz},
and KamLAND
\cite{Eguchi:2002dm}
experiments have introduced this era. 
In addition, the leptonic mixing angle $\theta_{13}$ has turned out to be small
by the results of the CHOOZ
experiment~\cite{Apollonio:1999ae,Apollonio:2002gd}.
Currently operating or planned conventional beam
experiments~\cite{Nakamura:2001tr,Paolone:2001am,Ereditato:2001an}
and planned superbeam experiments~\cite{Itow:2001ee,Ayres:2002nm,Beavis:2002ye}
will complement this information by further measurements of
$\theta_{13}$, the neutrino mass hierarchy, and the leptonic CP
violating phase $\deltacp$. Finally, future neutrino factories
(for a summary, see~\cite{Apollonio:2002en})
could find $\theta_{13}>0$ and test leptonic CP violation even below $\sin^2 2
\theta_{13} \lesssim 10^{-4}$~\cite{Huber:2003ak}.

Because of the high precision of these future experiments, it is
important to study the impact of matter density
effects on neutrino oscillations, since they may
affect the different measurements in a substantial way. It has for a
long time been known that neutrino oscillations are influenced by the
presence of matter~\cite{Wolfenstein:1978ue,Mikheev:1985gs,Mikheev:1986wj},
and therefore, matter density effects on neutrino oscillations in the
Earth have been thoroughly investigated in different contexts and with a
variety of models (see, for example, \Ref~\cite{Jacobsson:2001zk} and
references therein). In most of these analyses, the Preliminary Reference
Earth Model (PREM) matter density profile~\cite{Dziewonski:1981xy} has
been used, which has been obtained from geophysical seismic wave measurements.
However, the knowledge on the PREM matter density profile along a certain
baseline through the Earth's mantle is limited from geophysics to
about 5\% precision \cite{Pana,Geller:2001ix}, which means that matter
density uncertainties can affect the precision measurements of the
neutrino oscillation parameters, such as the leptonic mixing parameters
($\theta_{12}$, $\theta_{13}$, $\theta_{23}$, and $\deltacp$) as well
as the neutrino mass squared differences ($\Delta m_{21}^2$ and
$\Delta m_{31}^2$).

In this work, we will mainly focus on the effects of the matter density
uncertainties on the future measurements
of $\theta_{13}$, ${\rm sgn}(\Delta m_{31}^2)$, and $\deltacp$, since
these measurements are the most interesting ones for future long
baseline experiments. Especially, neutrino factories
could be affected by matter densities for two
reasons. First, they will be operating in the statistics
dominated regime with a very high precision. Second, in comparison
with superbeams, they are often proposed with very long baselines $L \gtrsim
700 \, \mathrm{km}$ with energy spectra covering the matter density
resonance in the Earth's mantle.
Thus, they will be strongly influenced by matter density effects
themselves (see, for example,
\Refs~\cite{Arafune:1997hd,Minakata:1998bf,Lipari:1999wy,Narayan:1999ck,Freund:1999gy,Campanelli:2000wi,Mocioiu:2000st,Ota:2000hf,Freund:2000ti,Miura:2001pi,Brahmachari:2003bk})
and also by matter density uncertainty effects
\cite{Koike:1998hy,Burguet-Castell:2001ez,Miura:2001pi2,Shan:2001br,Jacobsson:2001zk,Fogli:2001tm,Shan:2002px,Huber:2002mx,Jacobsson:2002nb,Ota:2002fu,Huber:2002rs,Huber:2003ak,Shan:2003vh,Huber:2003pm,Kozlovskaya:2003kk}.

The paper is organized as follows. In \Sec~\ref{sec:model}, we classify the
in the literature existing models for matter density uncertainties into 
three different categories. Next, in \Sec~\ref{sec:exp_des_sim}, we
give the experimental description of the neutrino factory setup that
we have been using for our analysis as well as we describe how the
simulations have been carried out. Then, in \Sec~\ref{sec:qualitative}, we
present a qualitative discussion of the matter density uncertainty
effects in the neutrino oscillation transition probabilities for the
appearance channels $\nu_e \to \nu_\mu$, $\nu_\mu \to \nu_e$,
$\bar\nu_e \to \bar\nu_\mu$, and $\bar\nu_\mu \to \bar\nu_e$.
In \Sec~\ref{sec:results}, we study the impact of matter density
uncertainties on the most interesting future measurements at a
neutrino factory. Furthermore, we show examples of such
measurements.
Finally, in \Sec~\ref{sec:s&c}, we present a summary of the obtained
results as well as our conclusions.

\section{Modeling matter density uncertainties}
\label{sec:model}

In most experimental simulations, the matter density is assumed to be
constant along the baseline: First, taking
$N$ matter density layers instead of one essentially increases the
computational effort by a factor of $N$. Second, the differences
in the results are, especially for not too long baselines, minor
compared to the increased effort and the effects of correlations and
degeneracies. Therefore, the matter density profile effect, \ie , the
difference between a constant and a realistic matter density profile,
is often not taken into account. However, for a fit of the data for a future
long-baseline experiment it should not be a problem to
incorporate it at least at the level of the PREM matter density profile.

In this work, we are interested in the uncertainties of such a
profile, which are coming from geophysics. These uncertainties have
different effects than the matter density profile itself and they
would essentially act as systematical errors in the
measurement. However, the matter 
density profile can, up to a certain level, also be absorbed as such
an uncertainty into the constant matter density~\cite{Ota:2002fu}. In
the following sections, we will therefore mainly study the consequences of
matter density uncertainties as perturbance to a constant matter density
profile, which should for specific experiments, of course, be replaced
by the profile, which is as close to the realistic profile as
possible. 

\begin{figure}[t!]
\begin{center}
\includegraphics[width=9cm]{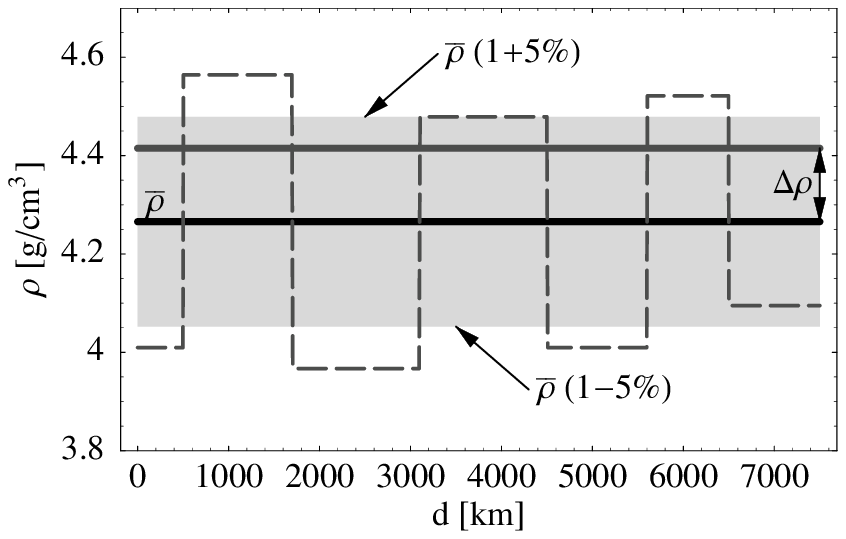}
\end{center}
\mycaption{\label{fig:model} The model of the average matter density as
  a measured quantity, illustrated for a baseline of $L = 7 \, 500 \,
  \mathrm{km}$ and a matter density uncertainty of $\Delta \rho^* =
  5\% \cdot \bar{\rho}$. The deviation from the average matter density
  $\bar{\rho}$, \ie, the quantity $\Delta \rho$, is
  measured by the experiment within the gray-shaded area. This area is
  determined by $\Delta \rho^* = \bar{\rho} \, (1 \pm 5 \%)$ at the $1
  \sigma$ confidence level with Gaussian statistics. A possible
  profile from the random fluctuations model in
  \Ref~\cite{Jacobsson:2001zk} is shown as the dashed curve.}
\end{figure}

In the literature, there exist, in principle, three different classes
of models for matter density uncertainties:
\begin{description}
\item[Statistical models to study the effects of realistic
  fluctuations.] These models are averaging over the effects of many
  profiles with similar properties to be as realistic as
  possible. Then, the average characteristic effects of matter density
  uncertainties on the neutrino oscillation transition probabilities
  or other quantities of interest are studied. Examples are the path
  integral method to average over many profiles in
  \Refs~\cite{Shan:2001br,Shan:2002px,Shan:2003vh} and the random
  fluctuations model in
  \Refs~\cite{Jacobsson:2001zk,Jacobsson:2002nb}. In this work, we use
  the latter one, which demonstrates the dependence on the length
  scale $\lambda$ and the amplitude $\Delta \rho$ of the
  fluctuations. In this approach, the actual value of the matter
  density $\rho$ is
  fluctuating around the average matter density $\bar{\rho}$ with the length
  scale $\lambda$ and the amplitude $\Delta \rho$ following some
  (truncated) Gaussian distributions with standard deviations
  $\sigma_\lambda$ and $\sigma_{\Delta \rho}$, as it is illustrated in
  \figu{model} with the dashed curve. From the discrepancies of
  realistic geophysics profile reconstructions~\cite{Pana}, one can
  estimate a set of standard parameters for these four parameters,
  which we will use if not otherwise stated: $\lambda \sim 2 \, 000
  \, \mathrm{km}$, $\sigma_{\lambda} \sim 3/4 \, \lambda$, $\Delta
  \rho_0 \sim 3 \% \, \bar{\rho}$, and $\sigma_{\Delta \rho} \sim 1/3
  \, \Delta \rho$.
\item[Simulation models assuming the matter density to be unknown.] The matter
  density profile is assumed to be unknown within certain limits in
  order to simulate the matter density uncertainty effects on neutrino
  oscillation parameter measurements. In these models, the unknown
  matter density profile (up to a certain degree) implies new
  parameters to be determined by the measurements,
  which, of course, are as well correlated amongst themselves as they
  are correlated with the neutrino oscillation parameters to be
  measured. Examples include the Fourier expansion method in
  \Refs~\cite{Fogli:2001tm,Ota:2000hf,Ota:2002fu} and models assuming
  an uncertainty for the average matter density in
  \Refs~\cite{Burguet-Castell:2001ez,Huber:2002mx,Huber:2002rs,Huber:2003ak,Huber:2003pm}.
  In \Ref~\cite{Ota:2002fu}, it is shown that the most interesting
  effects of matter density uncertainties can be translated into an
  uncertainty (or a shift) of the average matter density along the baseline. In
  fact, the matter density profile effect can also be incorporated
  into the model by this 
  approach, where an uncertainty of $5\%$ on the average matter density is
  quite a safe choice for not too long baselines. The ``model of the
  measured mean matter density'', which is illustrated in \figu{model}, uses
  the average matter density from the PREM matter density profile and
  allows it to vary within some range $\Delta \rho^*$. It is chosen to
  be $\Delta \rho^*=5\% \cdot \bar{\rho}$ 
  in \Refs~\cite{Huber:2002mx,Huber:2002rs,Huber:2003ak,Huber:2003pm}
  using this model in order to safely cover the effects of the matter
  density uncertainties and even the profile effect for not too long
  baselines. This means that the average matter density is measured
  by the experiment within $5\%$ externally imposed precision.
  {}From \figu{model} it should be clear that the random
  fluctuations can also be simulated with this model, since they are
  partially averaging out. However, it can be shown that for too large
  interference effects among the different matter density layers, such
  as for baselines $L \gtrsim 5 \, 000 \, \mathrm{km}$, the shape of
  the spectrum changes too much to simply ignore the matter density profile
  effect. Thus, experiments must, in their data analyses, take into
  account the profile effects for such long baselines even within this
  model.
\item[Models for a specific baseline.] For a specific baseline, one
  can try to obtain further geophysical information in order to
  incorporate the matter density profile as good as possible (see, for
  example, \Ref~\cite{Kozlovskaya:2003kk}). The deviations from the
  PREM matter density profile can in this case be rather
  large. However, the matter density uncertainties of such a profile
  are drastically reduced.
\end{description} 
\begin{figure}[t!]
\begin{center}
\includegraphics[width=8cm]{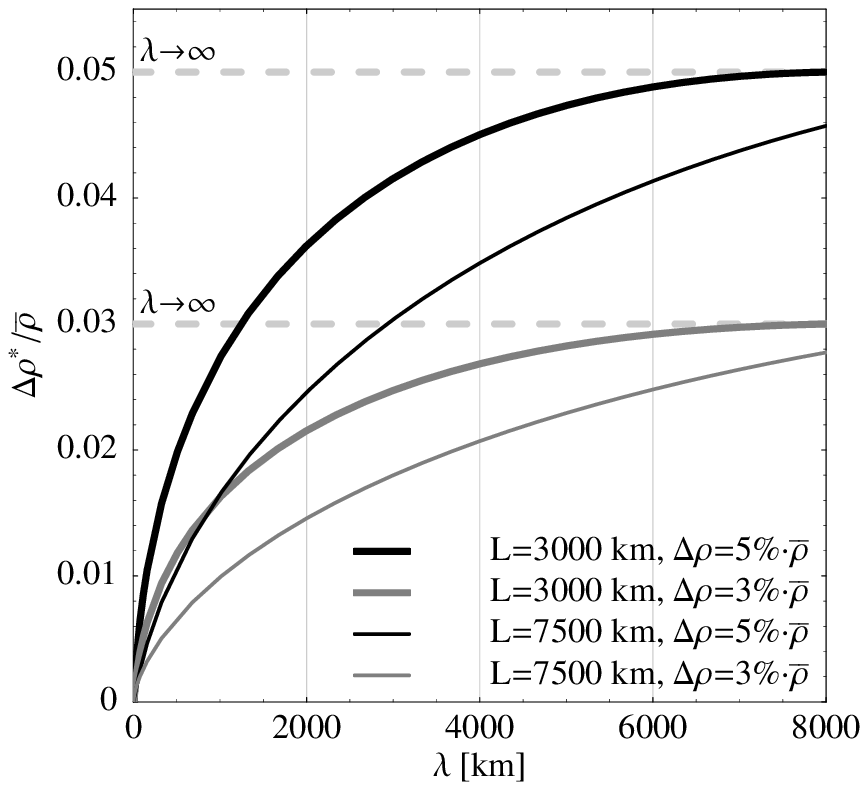}
\end{center}
\mycaption{\label{fig:mapping} The mapping between the length scale
  $\lambda$ of the random fluctuations model and the quantity $\Delta
  \rho^*/\bar{\rho}$ of the measured mean model for different
  fluctuation amplitudes and baselines. For the other parameters from
  the random fluctuations model in \Ref~\cite{Jacobsson:2001zk} the
  standard parameters therein were used: $\sigma_\lambda = 3/4 \,
  \lambda$ and $\sigma_{\Delta \rho} = 1/3 \, \Delta \rho$. The
  mapping is performed by minimizing the $\chi^2$-function for the
  \NuFactII\ experiment (\cf, \Sec~\ref{sec:exp_des_sim}) between the
  energy spectrum of 
  a random matter density profile and an average matter density
  profile with the density $\bar{\rho} + \Delta \rho^*$ with respect
  to the matter density shift $\Delta \rho^*$. It turns out that the minimal
  $\Delta \chi^2$ is, for short enough baselines, much smaller than
  unity, which means that the random matter density profile can be
  approximated by an average matter density profile
  to a very good approximation. The shown functions are computed by
  averaging the shifts $\Delta \rho^*$ over $2 \, 000$ such random
  profiles and using interpolation.}
\end{figure}
{From} the above discussion, it should be obvious that the models within each
class are rather similar to each other, although they may be using
different approaches and are aiming at different goals. However, one can
also transfer knowledge from the statistical models in
order to use it for a better estimation of the parameters in the
simulation models with unknown matter density parameters. For example, the
random fluctuations model (which simulates the effects of the
geophysical matter density uncertainties for different length scales
and amplitudes to be obtained from geophysics maps) can be used to
estimate the precision of the matter density within which it is
assumed to be unknown. Let us assume that we can identify the
fluctuations to have a certain length scale $\lambda$ and amplitude
$\Delta \rho$, what is the deviation of the average matter density $\Delta
\rho^*$, which needs to be used in the simulations?

As it is noted in \Ref~\cite{Ota:2002fu}, the energy spectrum with the
fluctuations can, for short enough baselines, be fit to the one with a
constant, but different, average matter density. The average difference between
the original and fit average matter densities can then be taken as a
measure for the allowed error on the average matter density $\Delta
\rho^*$. In \figu{mapping}, this mapping is performed as a function of
the length scale $\lambda$ for two different baselines and two
different amplitudes of the fluctuations. In this figure, it can
clearly be seen that the fluctuations partly average out for short length
scales. However, for length scales $\lambda \gg L$ there is no
difference between the random fluctuations and the measured average
models, since they are identical in this limit. It is also demonstrated
that in many cases a somewhat smaller value of $\Delta \rho^*$ than
$5\%$ should be sufficient for a fluctuation amplitude of $\Delta \rho^*
= 5\% \cdot \bar\rho$. However, in this case, the matter density
profile effect is not taken into account and it should be kept in mind
that we are only comparing average statements.

Closing this argumentation, the different length scales $\lambda$ also
correspond to the length scales of certain Fourier modes, which means
that fluctuations at a certain length scale leads to an uncertainty of
certain Fourier coefficients. These can by their correlation
(see \Ref~\cite{Ota:2002fu}) be translated into an uncertainty of the
leading Fourier coefficient. This means that the strength of this
correlation between the higher order Fourier coefficients and the
leading ones is another qualitative interpretation of
\figu{mapping}. We will therefore further on focus on the effects of 
the average matter density as a measured quantity in a full
statistical analysis, since this should simulate the matter density
uncertainties very well and it is suitable for more sophisticated
statistical simulations including correlations and degeneracies
(\cf, \Sec~\ref{sec:exp_des_sim}).

\section{Description of experiment and simulation}
\label{sec:exp_des_sim}

Although superbeams are often proposed with quite long baselines, they
are either operating in the statistics dominated regime
(first-generation experiments) or are limited by the backgrounds and
their uncertainties (superbeam upgrades). Therefore, the class of
experiments for which matter density uncertainties are most interesting
is neutrino factories: First, they usually have long enough baselines,
which means that matter effects substantially influence the appearance
rates. Second, their precision is very good, which means that matter
density uncertainties become a major impact factor. In order to demonstrate
the effects of matter density uncertainties, we use the advanced stage
neutrino factory setup \NuFactII\ from \Ref~\cite{Huber:2002mx}
with a muon energy of $50 \, \mathrm{GeV}$. This setup has a target power of
$4 \, \mathrm{MW}$, corresponding to $5.3 \cdot 10^{20}$ useful muon
decays per year. In addition, we assume eight years of total running
time, four of these with a neutrino beam and the other four with an
antineutrino beam. For the detector, we simulate a magnetized iron
detector with an overall fiducial mass of $50 \, \mathrm{kt}$. The
baseline length is assumed to be $L= 3 \, 000 \, \mathrm{km}$, which
has turned out to be reasonable for CP measurements. We use the
analysis technique, which is described in the Appendices A, B, and C
of \Ref~\cite{Huber:2002mx}, including the beam and detector
simulations. As described above, we use the average matter density
$\bar{\rho} = 3.5 \, \mathrm{g/cm^3}$, which corresponds to this baseline,
and we allow an uncertainty of $\Delta \rho^* = 5 \% \cdot
\bar\rho$. Later, we will discuss the effect of this 
uncertainty in greater detail. This means that the matter density is
treated as all the other neutrino oscillation parameters to be measured by the
neutrino factory. However, the average matter density is assumed to be
known from external measurements within some uncertainty, and therefore,
the matter density is considered as an external input coming from
geophysics (see also \figu{model}). This treatment does not only take
into account that the matter density is not precisely known, but it also
allows correlations between the neutrino oscillation parameters and the matter
density (\cf, \Ref~\cite{Ota:2002fu}). Since we include all
relevant appearance and disappearance channels in our analysis, the
only parameters which are measured much better with a different class
of experiments are the solar parameters. Hence, we use a similar
technique to include the external input from the solar parameters and we
assume that their product, which the neutrino factories are sensitive to,
is measured with $15\%$ precision by the KamLAND experiment by
the time the neutrino factory experiment would take
place~\cite{BARGER,Gonzalez-Garcia:2001zy}.

For the analysis of the quantities of interest we include, where
applicable, systematics, multi-parameter correlations, and degeneracies
as it is described in \Ref~\cite{Huber:2002mx} for the individual
measurements.
To summarize, the treatment of systematics is done in a
straightforward way by minimizing the $\chi^2$-function over the
auxiliary systematic parameters. Multi-parameter correlations are
included by projecting the $n$-dimensional fit manifold onto the axis
of interest. This approach always needs to be applied if one wants to
know the precision of an individual parameter, since the other parameters (such
as the CP phase) cannot be determined better by the experiment itself
and all possible points within the fit manifold are equally
allowed.\footnote{If the other parameters (that the quantity of interest is
  correlated with) were better determined by preceding measurements, then
  they should be either included as external input or, even better, by
  using the $\chi^2$-function of the relevant experiments. This does
  especially not apply to the CP phase $\deltacp$.} Alternatively,
one could, as a 
matter of taste, also project onto the plane of two parameters in
order to have the information less condensed, but this approach makes
it harder to discuss the dependence on the other parameters, such as the
true parameter values. As an important difference to the analysis of
existing experiments, the reference data set (event rate spectrum) of
future experiments is generated with a given set of true parameter
values assumed to be realized by Nature and it is not provided by a
measurement. In many cases, the results strongly depend on the true
parameter values, which can vary within their currently allowed
ranges. It is therefore our philosophy to condense the information as
much as possible in order to be able to discuss the true parameter
value dependencies, which are especially relevant to systematically
assess the potential of future experiments.

As far as degeneracies are concerned, we know the $(\deltacp,
\theta_{13})$~\cite{Burguet-Castell:2001ez}, $\mathrm{sgn}(\Delta
m_{31}^2)$~\cite{Minakata:2001qm}, and
$(\theta_{23},\pi/2-\theta_{23})$~\cite{Fogli:1996pv} degeneracies,
which lead to an overall ``eight-fold''
degeneracy~\cite{Barger:2001yr}. Since the current best-fit value for
the atmospheric mixing angle is $\theta_{23}=\pi/4$, we only consider the
first two degeneracies. The definition of the quantity of interest
quite often already includes the way the degeneracies have to be
treated. For example, defining the sensitivity to $\stheta$ as
the largest value of $\stheta$, which cannot be distinguished from
$\stheta=0$ (at the chosen confidence level) already implies that {\em
any} value of $\stheta$ fitting $\stheta=0$ of {\em any} degeneracy
is below this sensitivity limit. Thus, for $\stheta$ above this
sensitivity limit it is guaranteed that we will find $\stheta>0$
with the proposed experiment at the chosen confidence level. 
However, in some cases, including
degeneracies is not that straightforward. For example, for the
precision of the measurement of $\stheta$ we do not find a simple way
to include them, and therefore, we only show the results for the best-fit
manifold. Similar to the correlations, an existing experiment would
analyze its data in a different way. Such as it was performed for a
long time with the solar parameters, different isolated islands in the
parameter space would be allowed and subsequently reduced by the
future experiments (or the combination of some experiments). However, it is
the goal of the analyses of future experiments to keep the number
(determined by the degeneracies) and the extensions (determined by the
correlations) of these islands as small as possible by optimizing the
experiments before they are built. Therefore, it is reasonable to
condense the information as much as possible in order to discuss the
dependencies on the true parameter values.

For the neutrino oscillation parameters, we choose (if not otherwise stated)
the best-fit values $\Delta m_{31}^2 = 3.0 \cdot 10^{-3} \,
\mathrm{eV}^2$, $\sin^2 2 \theta_{23} =
1.0$~\cite{Gonzalez-Garcia:2002mu}, $\Delta m_{21}^2 = 7.0 \cdot 
10^{-5} \, \mathrm{eV}^2$, $\sin^2 2 \theta_{12} = 0.8$ (see, for
example, \Ref~\cite{Maltoni:2002aw}), as well as we only allow values
of $\sin^2 2 \theta_{13}$ below the CHOOZ bound and we assume a normal
mass hierarchy, \ie, ${\rm sgn}(\Delta m_{31}^2) = 1$. Because of the
symmetric operation of the neutrino factory, results would not differ
much qualitatively for an inverted mass hierarchy. In general, we do
not make any special assumptions about the true value of the CP phase
$\delta_{\mathrm{CP}}$. However, in some figures, we will choose
certain values for illustrative purposes, since studying the
dependence on the CP phase is not the subject of this work.

\section{Qualitative discussion of the appearance channels}
\label{sec:qualitative}

The goal of this section is to arrive at a qualitative understanding of 
the effects of matter density uncertainties. However, we will see that
there are quite many factors involved in this issue, which we will
discuss in the following in somewhat greater detail. At the end of
this section, we then will conclude with some very basic qualitative
observations.

For long-baseline experiments, the appearance probability
$P_{\mathrm{app}} = P_{e \mu}$, $P_{\mu e}$, $P_{\bar{e} \bar{\mu}}$,
or $P_{\bar{\mu} \bar{e}}$ in matter can be expanded in the small mass
hierarchy parameter $\alpha \equiv \Delta m_{21}^2/\Delta m_{31}^2$
and the small leptonic mixing angle $\theta_{13}$ up to second order
in these parameters as~\cite{CERVERA,FHL,FREUND}:
\begin{eqnarray}
P_{\mathrm{app}} &\equiv& P_1 \pm P_2 + P_3 + P_4 \nonumber\\
& \simeq & \sin^2 2\theta_{13} \, \, \sin^2
\theta_{23} \, \frac{\sin^2[(1- \hat{A}){\hat\Delta}]}{(1-\hat{A})^2}
\nonumber\\
&\pm& \alpha \, \sin 2\theta_{13} \, \, \sin
2\theta_{12} \, \cos\theta_{13} \, \sin 2\theta_{23} \, \sin
\delta_{\mathrm{CP}} \, \sin \hat\Delta \,
\frac{\sin(\hat{A}{\hat\Delta})}{\hat{A}} \,
\frac{\sin[(1-\hat{A}){\hat\Delta}]}{(1-\hat{A})} \nonumber\\
&+& \alpha \, \sin 2\theta_{13} \, \, \sin
2\theta_{12} \, \cos\theta_{13} \, \sin 2\theta_{23} \, \cos
\delta_{\mathrm{CP}} \, \cos \hat\Delta \,
\frac{\sin(\hat{A}{\hat\Delta})}{\hat{A}} \,
\frac{\sin[(1-\hat{A}){\hat\Delta}]} {(1-\hat{A})} \nonumber\\
&+& \alpha^2 \, \, \sin^2 2\theta_{12} \, \cos^2 \theta_{23} \,
\frac{\sin^2(\hat{A}{\hat\Delta})}{\hat{A}^2}.
\label{equ:PROBMATTER}
\end{eqnarray}
Here $\hat\Delta \equiv \Delta m_{31}^2 L/(4 E)$ and $\hat{A} \equiv \pm
(2 \sqrt{2} G_F n_e E)/\Delta m_{31}^2$ with $G_F$ being the Fermi coupling
constant, $L$ the baseline length, $E$ the neutrino energy, and $n_e$
the (constant) electron density in matter. This
expansion is qualitatively valid independently of the channel used,
but the signs before the second term and in the definition of the
parameter $\hat{A}$ depend on the direction $\nu_e \leftrightarrow
\nu_\mu$ and if one is using neutrinos or antineutrinos.
The sign before the second term is positive for $\nu_e \to \nu_\mu$
and neutrinos or $\nu_\mu \to \nu_e$ and antineutrinos and it is
negative for $\nu_e \to \nu_\mu$ and antineutrinos  or $\nu_\mu \to
\nu_e$ and neutrinos. The sign in $\hat{A}$ is positive for neutrinos
and negative for antineutrinos. Using parameter values for $\alpha$
within the KamLAND-allowed range, it turns out that for long-baseline
experiments all four terms in \equ{PROBMATTER} contribute to the appearance
channels with similar orders of magnitude. However, the relative
weight of these four terms is determined by the true values of the
parameters $\alpha$ and $\theta_{13}$.

In \equ{PROBMATTER}, the matter potential enters via $\hat{A}$ and the
matter density resonance condition can be expressed as $\hat{A} \simeq
1$. At the resonance, the factor $\sin [ ( 1-\hat{A}) \hat\Delta]/(1-
\hat{A})$ becomes maximal, which means that the first three terms of
\equ{PROBMATTER} are strongly influenced by the resonance
condition. Especially, the quadratic dependency in the first term of
\equ{PROBMATTER} makes it peak much stronger at the resonance than the
second and third terms, and therefore suggests that it is most
sensitive to a change in any of the parameters in $\hat{A}$.
The effect of such a parameter change would also create a shift of the
position of the resonance in the neutrino energy spectrum. This is,
for example, often used for the sign of $\ldm$-measurement, since
$\hat{A}$ changes sign for the inverted mass hierarchy, producing a
very distinctive solution. In this case, the resonance is essentially
shifted towards negative neutrino energies, which is, of 
course, an unphysical solution. Since the first term of \equ{PROBMATTER} is
proportional to $\stheta$, mass hierarchy measurements are therefore
favored for large values of $\stheta$ (and small values of $\alpha$ in
order to keep the other terms small, which cause correlation and
degeneracy problems for this measurement). Another parameter, which
enters in $\hat{A}$, is the electron density $n_e$ representing the
matter density $\rho$.\footnote{The electron density $n_e$ is related to the
matter density $\rho$ as $n_e = \frac{Y_e}{m_N} \rho$, where $Y_e$ is the
average number of electrons per nucleon (in the Earth: $Y_e \simeq
\frac{1}{2}$) and $m_N \simeq 940 \, {\rm MeV}$ is the nucleon mass.}
Similar to the sign of $\ldm$, we therefore expect
that the neutrino energy spectrum is especially deformed by the first term
in \equ{PROBMATTER} for a shift of the matter density.

In order to demonstrate the above, let us investigate the impact of a
perturbation (or shift) $\Delta \hat{A}$ on the parameter $\hat{A}$ 
somewhat closer. This shift $\Delta \hat{A}$
corresponds to the amplitude of the matter density uncertainty $\Delta
\rho$. One can write the absolute shift of the appearance probability
$P_{\rm app}$ up to first order in the relative shift of the parameter
$\hat{A}$ as:\footnote{A comparison with an exact numerical
  calculation suggests that this first order approximation is
  sufficient for our purposes.}
\begin{eqnarray}
\Delta P_{\rm app} &\equiv& P_{\rm app}(\hat{A}+\Delta\hat{A}) - P_{\rm
  app}(\hat{A}) \nonumber\\
&=& \bigg[ - 2 \left( \hat{A} \hat\Delta \cot
  [(1-\hat{A}) \hat\Delta] - \frac{\hat{A}}{1-\hat{A}} \right) P_1 \nonumber\\
&+& \left( - \hat{A} \hat\Delta \cot [(1-\hat{A}) \hat\Delta] +
  \hat{A} \hat\Delta \cot (\hat{A} \hat\Delta) - \frac{1-2\hat{A}}{1-\hat{A}}
  \right) P_{2,3}
  \nonumber\\
&+& 2 \left( \hat{A} \hat\Delta \cot (\hat{A} \hat\Delta) - 1 \right)
  P_4 \bigg] \frac{\Delta \hat{A}}{\hat{A}} +
  \mathcal{O}\left((\Delta\hat{A}/\hat{A})^2\right),
\label{equ:DeltaP}
\end{eqnarray}
where $P_{2,3} \equiv \pm P_2 + P_3$.
\begin{figure}[t!]
\begin{center}
\includegraphics[width=8.5cm,angle=-90]{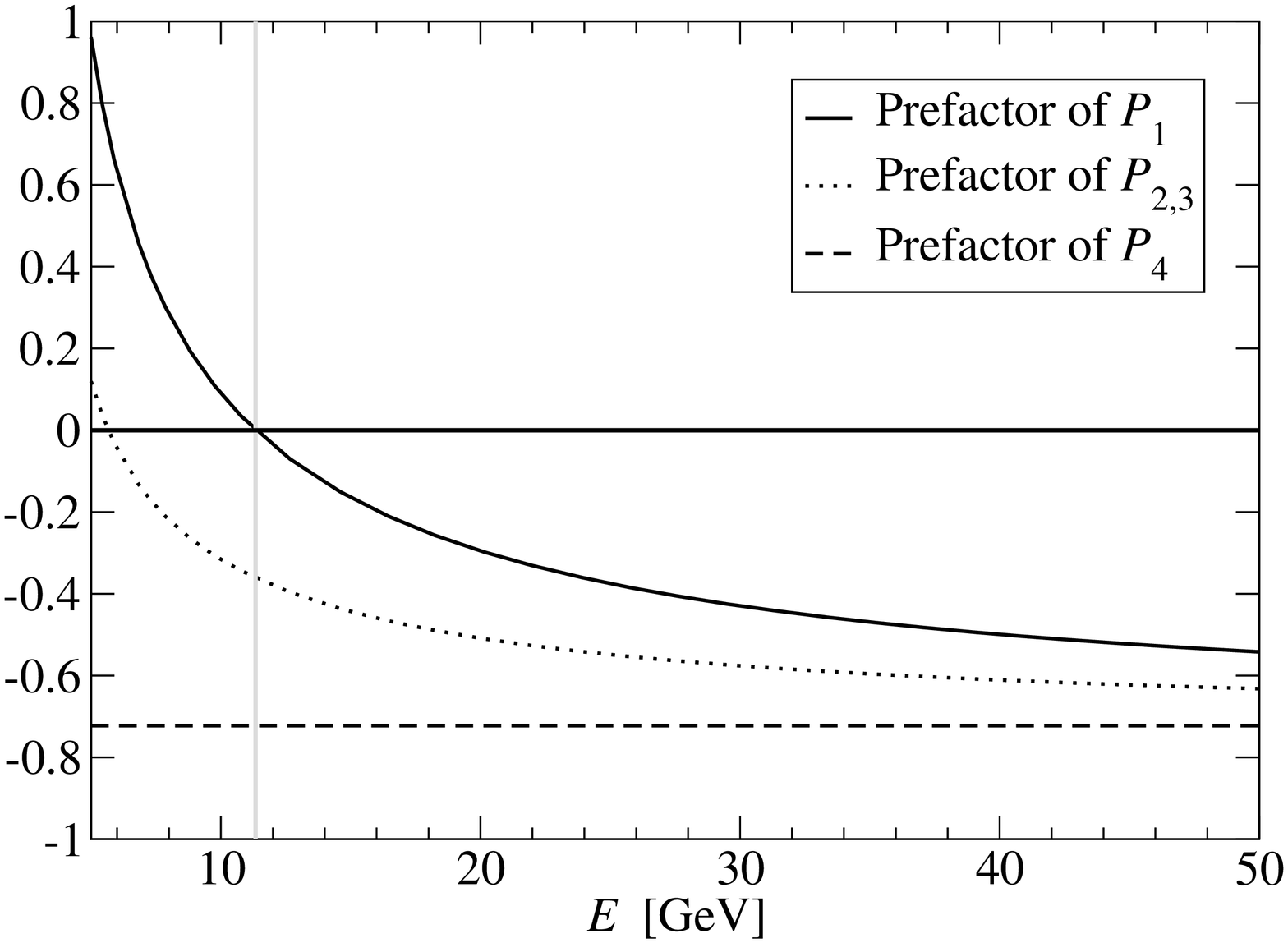}
\end{center}
\mycaption{\label{fig:factors} The prefactors of $P_1$, $P_{2,3}$, and
  $P_4$ in \equ{DeltaP} as functions of the neutrino energy $E$ ($5
  \, {\rm GeV} \leq E \leq 50 \, {\rm GeV}$) for
  the parameter values $\Delta m_{31}^2 = 3.0 \cdot 10^{-3} \, {\rm
  eV}^2$, $L = 3 \, 000 \, {\rm km}$, and $\rho = 3.5 \, {\rm
  g/cm^3}$. The vertical gray line marks the value of the resonant
  neutrino energy $E_{\rm res} \simeq 11 \, {\rm GeV}$.
}
\end{figure}
The absolute error of the appearance probability $P_{\rm app}$ is here
an appropriate representation for the statistical analysis, since the
change in the event rates and their impact on the $\chi^2$-value is
proportional to the absolute (and not relative) error. Note that all
prefactors appearing together with the $P_i$'s ($i =1,2,3,4$) in
\equ{DeltaP} are nearly of the same order of magnitude,
at least for neutrino energies $E \gtrsim 20 \, \mathrm{GeV}$, as it can
be seen from \figu{factors}. However, the orders of magnitude of the
different terms in \equ{DeltaP} can still be very different from each
other, since these terms are proportional to the absolute values of the
$P_i$'s defined in \equ{PROBMATTER}, which {\it per
se} could be very different. For example, for large values of
$\stheta$ the first term in \equ{PROBMATTER} is relatively large 
and would therefore also be strongly affected by matter density uncertainties.
In addition, all terms in \equ{DeltaP} are proportional to $\Delta
\hat{A}/\hat{A}$ to first order, which means that the absolute change
of one of the terms is obtained by multiplying the prefactor with the
relative change of the matter density and the corresponding $P_i$ itself.
In \figu{factors}, we observe that all prefactors are finite and not
singular at the resonant neutrino energy $E_{\rm res} \simeq 11 \,
{\rm GeV}$ and that the prefactor of $P_4$ is a constant with respect
to the neutrino energy $E$. Most interesting, the prefactor of $P_1$
changes sign at the resonance, which means that a small shift $\Delta
\hat{A}$ of the matter density close to the resonance will cause
\begin{itemize}
\item
a spectral distortion and a change of the normalization in the first
(and also somewhat in the second and third) term
\item
only a change of the normalization in the fourth term
\end{itemize}
of \equ{PROBMATTER}. Since the effects of the matter density shift can
be of equal size, this observation makes the analysis of matter
density uncertainties rather complicated. However, it should be noted
that especially the first term can also simulate spectral distortion effects in
addition to a change of the normalization, which means that it allows
more sophisticated correlations than the other terms.

As far as correlations are concerned, we observe that in
\equ{PROBMATTER} only the combinations
\begin{eqnarray}
C_1 & = &\sin 2 \theta_{13} \, \frac{\sin [ (1 - \hat{A}) \hat\Delta]}{1 -
\hat{A}} , \\
C_2 & = & \alpha \, \sin 2 \theta_{12} \,
\frac{\sin(\hat{A}{\hat\Delta})}{\hat{A}}
\end{eqnarray}
are present. In \equ{PROBMATTER}, $C_1^2$ appears in the first term,
$C_1 C_2$ in the second and third terms, and $C_2^2$ in the fourth
term. This means that a change in any parameter value in these terms
can be compensated by another one. Since $\stheta$ is the quantity of
interest to us, it will lead to correlations with the matter density
shift $\Delta \hat{A}$. However, the quantities $\alpha$ and $\sin 2
\theta_{12}$ are given by external measurements with lower precisions
than the matter density, which means that they can easily absorb a
change of the matter density from the correlation in the first term of
\equ{PROBMATTER}. Therefore, especially for large $\stheta$, where the
influence of matter density uncertainties to the first term in
\equ{PROBMATTER} is large (\cf, \equ{DeltaP}), the combination $C_1$
can have a major impact on measurements.
 
Another important ingredient in the analysis is statistics.
This means that the impact of the matter density uncertainties can be
suppressed in statistics dominated regions of the parameter space. Since some
measurements are limited by statistics, the matter density uncertainty
will not at all have a strong impact on these
measurements. For example, the $\stheta$-sensitivity limit describes
the ability for an experiment to establish $\stheta>0$. In this case,
the reference rate vector is computed with $\stheta = 0$ and the
statistics from the first three terms in \equ{PROBMATTER}, which are
only present for $\stheta>0$, limits the measurement. Thus, the
$\stheta$-sensitivity limit is constrained by the absolute size of these
terms, which cannot be compensated by the correlation with matter density
uncertainties because of the spectral dependence of the first
prefactor in \equ{DeltaP}. Since all of the quantities of interest
to us are suppressed by $\sin 2 \theta_{13}$ in \equ{PROBMATTER},
a similar argument can be used for small values of $\stheta$,
where most measurements are limited by the statistics in the
appearance channels and not by the matter density uncertainties.
In addition, matter density uncertainties will
turn out not to be relevant for many measurements with a strong
``binary nature'', \ie, measurements which only have two very
distinctive different answers. The ability to distinguish these two
answers often depends more on statistics than on the matter density
uncertainty, which cannot simulate the different solution. One such 
example is the mass hierarchy determination discussed above,
which could only be mimicked by matter density uncertainties 
larger than $100 \% \cdot \bar{\rho}$.

Putting all pieces together, we have found that especially $\stheta$
as a quantity of interest is highly correlated with the matter density
uncertainties. However, the effects of the matter density
uncertainties are proportional to the four terms in \equ{PROBMATTER}
themselves, which means that they are (for the quantities of interest)
suppressed by $\stheta$
itself. Therefore, we expect a large influence of matter density
uncertainties for large $\stheta$ and a small influence for small
$\stheta$ because of domination of statistics.
For the $\stheta$-precision measurement the correlation between
$\stheta$ and the matter density uncertainties should be limiting the
measurements for large values of $\stheta$. However, for CP measurements
the first term in \equ{PROBMATTER} acts as a background. The matter
density uncertainties should introduce an additional systematical
error for large values of $\stheta$, such as a background uncertainty.
In addition, $\deltacp$ is highly correlated with $\stheta$, which
again is correlated with $\Delta \hat{A}$, which suggests that it is
also indirectly affected via this correlations. However, since the
relative impact of the first term in \equ{PROBMATTER} decreases for
increasing $\alpha$, also the effects of the matter density
uncertainties should become smaller.

\section{The results for the different measurements}
\label{sec:results}

In this section, we systematically discuss the impact of matter
density uncertainties on the most interesting measurements at a
neutrino factory: $\stheta$, the sign of $\ldm$, and
$\deltacp$. Although many of these measurements have been individually
treated in previous works, the goal of this work is to specifically demonstrate
where in parameter space and for which measurements matter density
uncertainties 
are relevant and important. In addition, we will show, where applicable,
representative examples of such measurements as functions of the
amplitude of the matter density uncertainties. We choose
the ``model of the measured mean matter density'' with the values $\Delta
\rho^*=0$ (no matter density uncertainties), $\Delta
\rho^*=1\% \cdot \bar{\rho}$, $\Delta \rho^*=3\% \cdot \bar{\rho}$, $\Delta
\rho^*=5\% \cdot \bar{\rho}$ (the standard value, which we usually
propose as a conservative choice), and $\Delta \rho^*=10 \% \cdot
\bar{\rho}$ (the most pessimistic choice).

\subsubsection*{The sensitivity to $\mbox{\boldmath$\stheta$}$}

We define the sensitivity to $\stheta$ as the largest value of
$\stheta$, which fits the true value $\stheta=0$. As discussed
in the previous section, this choice naturally includes the treatment of
correlations and degeneracies, which means that any point in the
multi-dimensional parameter space, which fits the zero rate vector at
the chosen confidence level, will have a fit value of $\stheta$ below the
sensitivity limit. Thus, it is guaranteed that the experiment will
find $\stheta$ above the sensitivity limit at the chosen confidence
level. Any other definition of the sensitivity limit would be more
complicated and would involve a more sophisticated
definition of how to treat degeneracies. For example, one could show
the sensitive ranges for the different degeneracies separately.
Such a definition might be
sensible for existing experiments, but not to evaluate and compare the
potential of future experiments in order to optimize them using
condensed information.

As we qualitatively discussed in \Sec~\ref{sec:qualitative}, the
sensitivity to $\stheta$ is limited by the statistics
in the appearance channels. Therefore, the systematics 
introduced by the matter density uncertainties does not affect it
substantially. This can, for example, be seen from comparing \fig~9
with \fig~10 in \Ref~\cite{Huber:2002mx} for \NuFactII. In this
impact factor analysis, the matter density affects the
$\stheta$-precision measurements, but it does not at all affect the
$\stheta$-sensitivity limit.

\subsubsection*{The precision of $\mbox{\boldmath$\stheta$}$}

\begin{figure}[t!]
\begin{center}
\includegraphics[width=8cm]{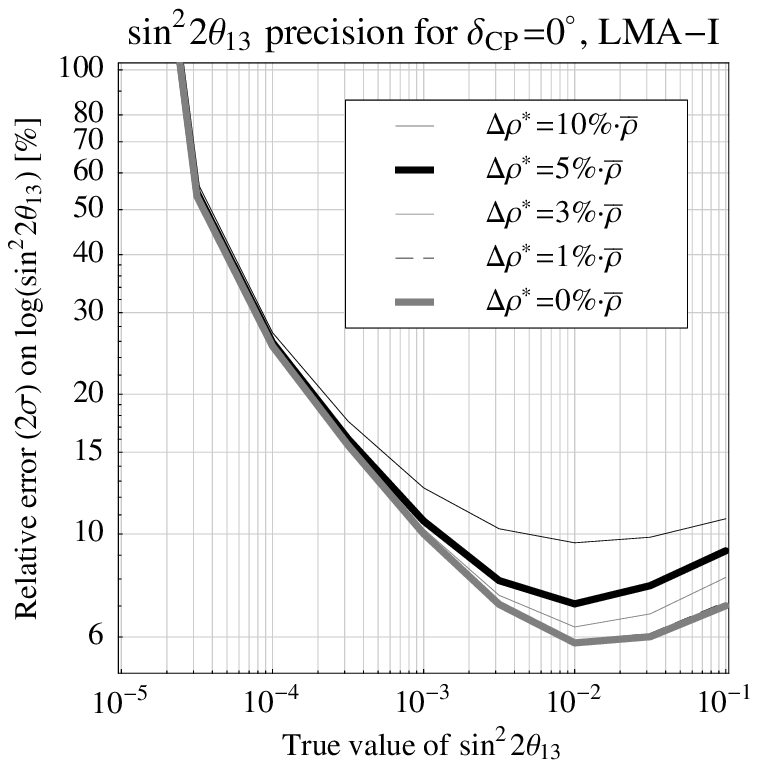}
\end{center}
\mycaption{\label{fig:th13matter} The precision of the measurement of
  $\stheta$ (full width, best-fit manifold only) for \NuFactII\ and
  the true value of $\deltacp=0$ as a function of the true value
  of $\stheta$ at the $2\sigma$ confidence level. The different curves
  correspond to different allowed matter density uncertainties $\Delta
  \rho^*$ as described in the plot legend, especially the thick curves
  correspond to no matter density uncertainty (light thick curve) and
  our standard uncertainty $\Delta \rho^* = 5 \% \cdot \bar{\rho}$
  (dark thick curve).}
\end{figure}

We define the precision of $\stheta$ as the full width error (at
the chosen confidence level) of the best-fit manifold projected onto
the $\stheta$-axis. This definition includes correlations, but it does
not include (disconnected) degeneracies, since a 
sensible definition of the $\stheta$-precision including disconnected
degeneracies is difficult. In \figu{th13matter}, we show the relative
error on $\mathrm{log}( \stheta )$ on a double-logarithmic scale. In
addition, we only show 
it for one true value of $\deltacp = 0$, since otherwise one could not
clearly see the effect of the matter density uncertainty. However, one
should keep in mind that this precision strongly depends on the true
value of $\deltacp$. A possible representation of this fact is to show
the band ranging over all true values of the CP phase, such as it is
done in \fig~13 of \Ref~\cite{Huber:2002mx}.

Figure~\ref{fig:th13matter} demonstrates the effects of the matter densities as
a function of their amplitude $\Delta \rho^*$ for the LMA-I solution,
where the dark thick curve corresponds to our standard choice and the
light thick curve to no matter density uncertainties, which is often used in
other studies. Except from the most extreme (and unrealistic) case of
$\Delta \rho^*=10 \% \cdot \bar\rho$, we do not find any significant
influence of the matter density uncertainties below $\stheta \lesssim
10^{-3}$. In this region, the $\stheta$-dependent terms in
\equ{PROBMATTER} are more determined by the statistics than the
matter density uncertainty. However, at the CHOOZ limit, the
matter density uncertainties lead to corrections at the percent level 
of the investigated quantity. It is also interesting to note
that for $\Delta \rho^* \lesssim 1 \% \cdot \bar{\rho}$ the matter
density uncertainties are irrelevant. Figure~\ref{fig:th13matter} is 
representative for different values of $\deltacp$ than
$\deltacp=0^\circ$. For example, if one drew bands using all values
of $\deltacp$ (such as in \fig~13 of \Ref~\cite{Huber:2002mx}),
the matter density uncertainties would lead to an upward shift of
these bands for large values of $\stheta$. A similar qualitative
behavior can also be expected for the LMA-II solution, although the
general performance with respect to $\stheta$ is worse because of the
stronger presence of correlations and degeneracies.

\subsubsection*{The sensitivity to the sign of $\mbox{\boldmath$\ldm$}$}

We define that there is a sensitivity to a certain sign of $\ldm$, if
there is no solution fitting the zero rate vector
(generated with this specific sign) with the opposite sign of $\ldm$
at the chosen confidence level. This definition implies that the
presence of the $\mathrm{sgn}(\ldm)$-degeneracy at the chosen confidence
level is the most important factor, which destroys the sign of the
$\ldm$-sensitivity. Performing the analysis, it turns out that the matter
density uncertainty hardly affects the appearance of the degenerate
solution. This means that matter density uncertainties are not
relevant for the sign of the $\ldm$-determination, at least within the
KamLAND-allowed range, even $\Delta \rho^* = 10 \% \cdot \bar{\rho}$ is
safe for \NuFactII . Given the definition of $\hat{A}$ at \equ{PROBMATTER},
a compensation of a different sign of $\ldm$ in $\hat{A}$ could only
be achieved by reversing the sign of $\rho$, \ie, negative matter
densities. However, in our analysis, such a reversion would require
matter density uncertainties of over $100 \% \cdot \bar{\rho}$ in order
to allow the unphysical assumption of antimatter in the Earth.

\subsubsection*{The sensitivity to {\em any} CP violation}

\begin{figure}[t!]
\begin{center}
\includegraphics[width=16cm]{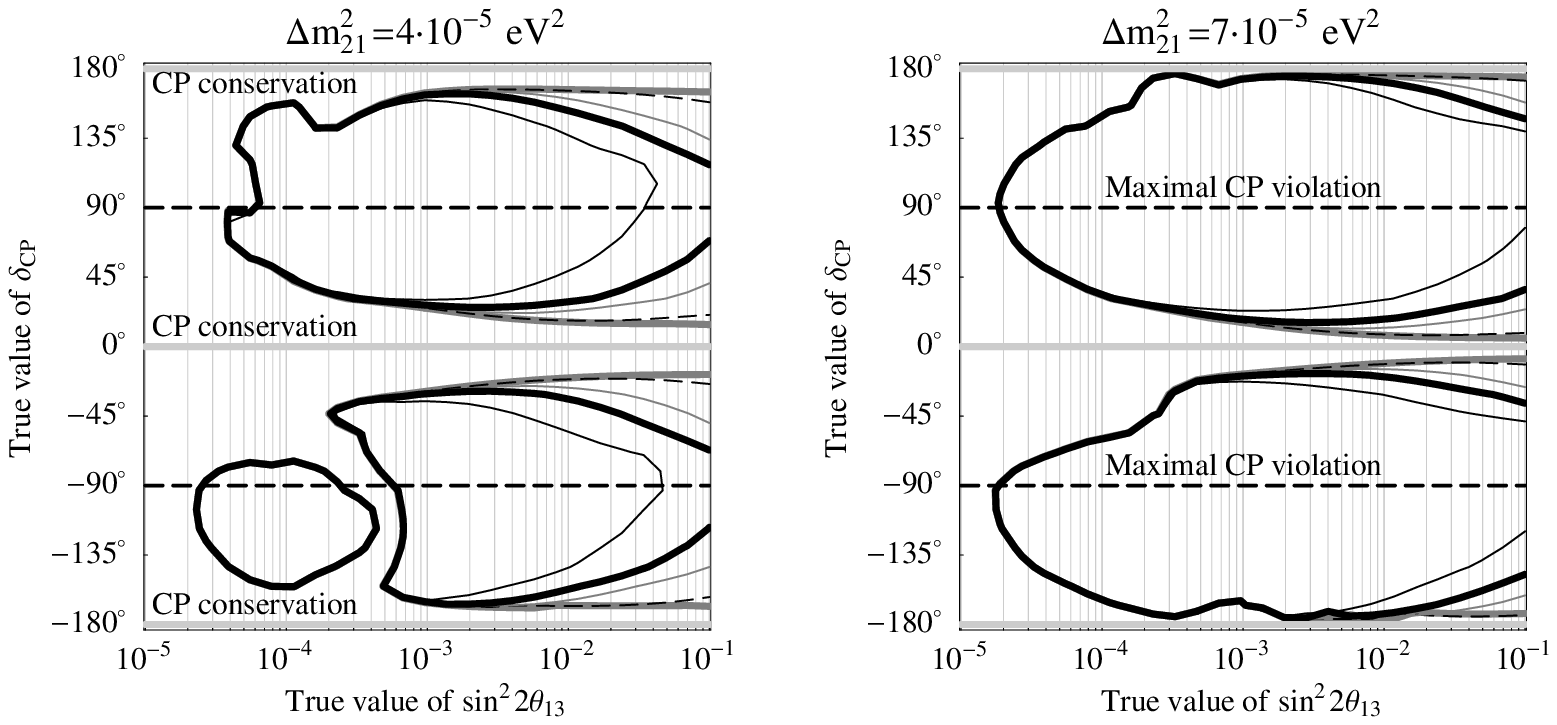}
\end{center}
\mycaption{\label{fig:cpviolmatter} The sensitivity to CP violation
  for \NuFactII\ and two different values of $\sdm$ as a function of
  the true values of $\stheta$ and $\deltacp$ at the $2\sigma$
  confidence level. The parameter values within the regions enclosed
  by the contours refer to the sensitivity to CP violation, \ie, the ability
  to distinguish the respective true value of $\deltacp$ given at the
  vertical axis from $\deltacp \in \{0, 180^\circ \}$ (CP
  conservation). The different curves correspond to different values of
  $\Delta \rho^*$ as given in the legend of \figu{th13matter}, especially
  the thick curves correspond to no matter density uncertainty (light
  thick curve) and our standard uncertainty $\Delta \rho^* = 5 \%
  \cdot \bar{\rho}$ (dark thick curve).}
\end{figure}

One can define the sensitivity to {\em any} CP violation as the
ability of an experiment to distinguish any given true CP violating
value $\deltacp \notin \{0,180^\circ\}$ from the CP conserving values
$\deltacp \in \{0,180^\circ\}$. This definition already implies the
treatment of degeneracies: any degenerate solution fitting $\deltacp
\in \{0,180^\circ\}$ would destroy the sensitivity to CP violation. A
special case of CP violation covered by this definition is the case of
{\em maximal} CP violation with the true value $\deltacp \in
\{-90^\circ,90^\circ\}$, which we will discuss in the next
subsection.

Figure~\ref{fig:cpviolmatter} shows the sensitivity to any CP
violation as a function of the true values of $\stheta$ and $\deltacp$
for several different values of the matter density uncertainty. The
left-hand side plot corresponds to the current lower limit of the
KamLAND-allowed range, whereas the right-hand side plot corresponds to
the LMA-I best-fit value. Again, the matter density uncertainties are
only relevant for $\stheta \gtrsim 10^{-3}$, as it can be clearly
seen in \figu{cpviolmatter}, and the
measurements are statistics dominated for $\stheta \lesssim 10^{-3}$.
Comparing the dark thick curves for our
standard value $\Delta \rho^* = 5\% \cdot \bar{\rho}$ between the left-
and right-hand side plots at the CHOOZ limit indicates that the relative
influence of the matter density uncertainties decreases with
increasing $\sdm$. In general, the CP performance becomes better for a
large hierarchy parameter $\alpha \equiv \sdm/\ldm$ in
\equ{PROBMATTER}, where the CP sensitive terms also become
large. Thus, the relative impact from the matter density uncertainties
in the first term of \equ{PROBMATTER} decreases for larger values of
$\alpha$. In addition, problems with degeneracies are reduced by
better statistics of the CP sensitive terms. Another important
conclusion from \figu{cpviolmatter} is that for $\Delta \rho^*
\lesssim 1 \% \cdot \bar{\rho}$ the matter density uncertainties are
again not relevant.

\subsubsection*{The sensitivity to {\em maximal} CP violation}

The special case of sensitivity to {\em maximal} CP violation
$\deltacp \in \{-90^\circ,90^\circ\}$ is represented by the horizontal
dashed lines in \figu{cpviolmatter}. This measurement has a much more
``binary nature'' than the sensitivity to a small CP violation, since
CP conservation and maximal CP violation produce rather distinctive
rate vectors for large values of $\stheta$, where matter density
uncertainties are most important. Since the sensitivity to maximal CP
violation is close to the optimum of all CP violation sensitivities,
it is often used as a representative for CP violation discussions. For
example, it is shown as a function of the true values of $\stheta$ and
$\sdm$ in \fig~17 of \Ref~\cite{Huber:2002mx}, a presentation, which
would not be possible for {\em any} CP violation.

It can be shown that matter density uncertainties are not important
for maximal CP violation measurements at the $2 \sigma$ confidence
level within the ($3 \sigma$) KamLAND-allowed range as long as $\Delta
\rho^* \lesssim 5 \% \cdot \bar{\rho}$. This is consistent with
\figu{cpviolmatter} along the horizontal dashed lines, since the
left-hand side plot represents the lower limit of the KamLAND-allowed
range and the relative influence of the matter density uncertainties
becomes smaller for larger values of $\sdm$. Only for very large matter density
uncertainties $\Delta \rho^* = 10 \% \cdot \bar{\rho}$, they have some
impact on the very lower end of the KamLAND-allowed range. However,
note that this statement does not mean that matter density
uncertainties in general do not affect the (maximal) CP violation
measurements. It depends on the used parameter values and it is supported
by the fact that the KamLAND experiment cut off the lower part of the
LMA solution.

\subsubsection*{The precision of $\mbox{\boldmath$\deltacp$}$}

\begin{figure}[t!]
\begin{center}
\includegraphics[width=8cm]{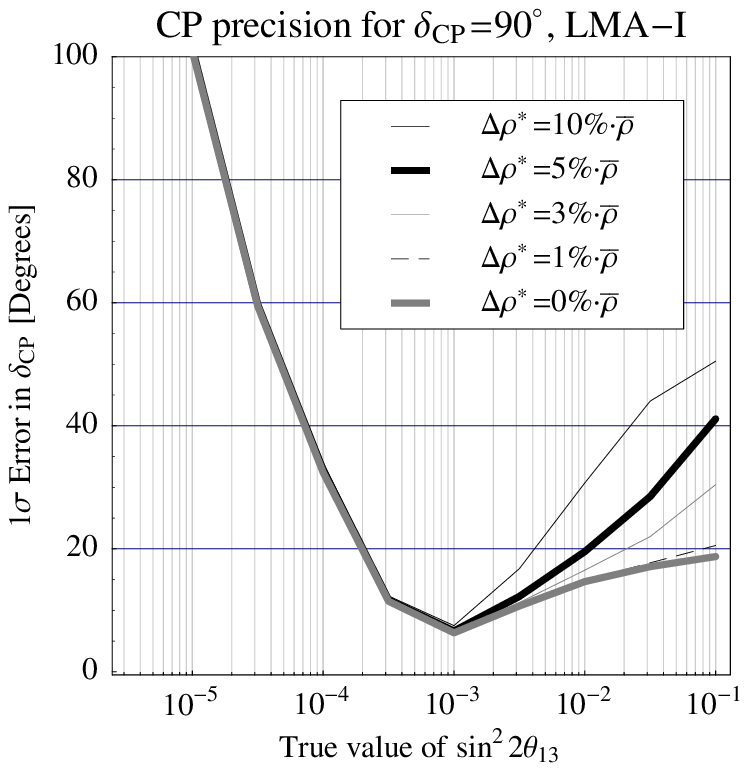}
\end{center}
\mycaption{\label{fig:cpprecmatter} The precision of the measurement
  of $\deltacp$ (full width) for \NuFactII\ and the true value of
  $\deltacp=90^\circ$ as a function of the true value of $\stheta$ at
  the $1\sigma$ confidence level. The different curves correspond to
  different allowed matter density uncertainties $\Delta \rho^*$ as
  described in the plot legend, especially the thick curves correspond to
  no matter density uncertainty (light thick curve) and our standard
  uncertainty $\Delta \rho^* = 5 \% \cdot \bar{\rho}$ (dark thick
  curve).}
\end{figure}

We define the precision of $\deltacp$ as the full width error of
$\deltacp$ on the ``CP circle''. For a given true value of $\deltacp$ 
any degenerate solution, which fits the true value, is included in this
precision. It is sometimes also called the ``coverage in
$\deltacp$''~\cite{Huber:2002mx}, because it describes how much of the
CP circle that fits the chosen true value of $\deltacp$. Thus, a precision
of $360^\circ$ (or a coverage of $100 \%$) corresponds to no
improvement of the knowledge on $\deltacp$. In comparison to the CP
violation sensitivity, the precision of $\deltacp$ does not differentiate any
special value of $\deltacp$, such as the CP conserving values $0$ and
$180^\circ$. It describes how much one could learn about
$\deltacp$ from a specific experiment. For example, even if an
experiment is not sensitive to CP violation, because $\deltacp$ is too
close to CP conservation, it may teach us something about $\deltacp$
and exclude certain regions of the CP circle. In this case, the
precision (coverage) would be smaller than $360^\circ$.

In \figu{cpprecmatter}, the precision of $\deltacp$ is shown as a
function of the true value of $\stheta$ for the true value
$\deltacp=90^\circ$ at the $1 \sigma$ confidence level. One could also
choose a different true value of $\deltacp$ or a different confidence
level, but the effects of the matter density uncertainties would
qualitatively look the same. However, degeneracies are hardly
present below the $1 \sigma$ confidence level. Therefore, the numbers
on the precision axis should be handled with care, since the $3
\sigma$ errors would be over-proportionally larger not following
simple Gaussian statistics. As for the case of
CP violation, matter density uncertainties become important for
$\stheta \gtrsim 10^{-3}$ and $\Delta \rho^* > 1 \% \cdot \bar{\rho}$
for the same reasons. However, since the CP precision measurement
includes all possible fit values of $\deltacp$ and it is not a binary
measurement, the effects of the matter density uncertainties can be
very large. At the CHOOZ limit, they can even affect the precision by a
factor of two or more.

\section{Summary and conclusions}
\label{sec:s&c}

\begin{figure}[t!]
\begin{center}
\includegraphics[width=8cm]{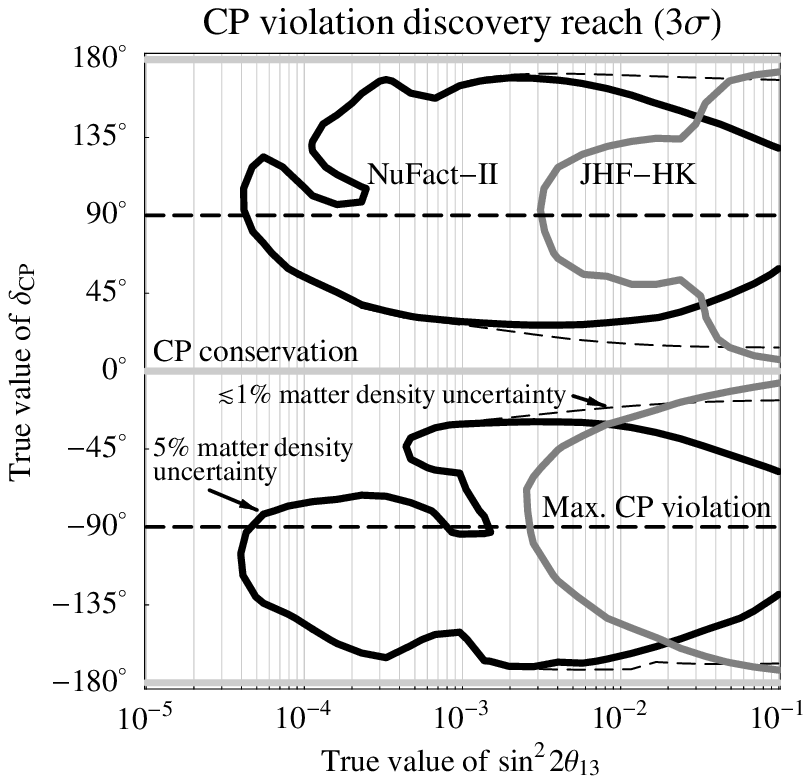}
\end{center}
\mycaption{\label{fig:cpviol3s} The $3 \sigma$ CP violation discovery
  reach as a function of the true values of $\stheta$ and $\deltacp$
  for \NuFactII\ and \JHFHK\ from \Ref~\cite{Huber:2002mx}. 
  The parameter values within the regions enclosed by the contours
  correspond to the sensitivity to CP violation, \ie, the ability
  to distinguish the respective true value of $\deltacp$ given at the
  vertical axis from $\deltacp \in \{0, 180^\circ \}$, \ie, CP
  conservation. The different curves correspond to different values of
  $\Delta \rho^*$ as given in the figure. For the \JHFHK\ experiment,
  matter effects (and matter density uncertainty effects) are strongly
  suppressed by the short baseline $L \simeq 295 \, \mathrm{km}$.
  For the \NuFactII\ experiment, degeneracies are much more present at
  the $3 \sigma$ confidence level than at the $2 \sigma$ confidence
  level, as it can be seen by comparison with \figu{cpviolmatter}
  (right-hand side plot).}
\end{figure}

Matter density uncertainties up to $5\%$ can be present on the PREM
matter density profile. It is well known that they can influence many
of the measurements at a future neutrino factory, such as the CP precision
measurements. In this work, we have first of all discussed different
approaches to model the matter density uncertainties and how they are
related to each other. We have concluded that using the average matter
density with an uncertainty of about $\Delta \rho^* = 3 \%$ - $5\%$ is
an appropriate conservative choice for a complete statistical neutrino factory
simulation. In this approach, the average matter density is measured
by the neutrino factory as an independent parameter within the
externally given precision $\Delta \rho^*$. It has several advantages:
\bi
\item
 It is fast enough for a complete statistical simulation including
 systematics, correlations, and degeneracies.
\item
 It allows correlations between the matter density and the neutrino oscillation
 parameters. 
\item
 It can use the external information from other statistical models or a
 specific baseline to improve the knowledge on the input parameter
 $\Delta \rho^*$.
\item
 It can even simulate the matter density profile effect for not too
 long baselines ($L \lesssim 5 \, 000 \, \mathrm{km}$).
\ei

With this model, we have systematically discussed how the matter
density uncertainties affect the most important measurements at a
large neutrino factory at a baseline of $L=3 \, 000 \,
\mathrm{km}$. We have found that some, but not all, measurements
suffer from the matter density uncertainties. In addition, they are
only relevant in certain regions of the parameter space. In detail, we
have found that
\begin{description}
\item[For the $\mbox{\boldmath$\stheta$}$-sensitivity] matter density
  uncertainties are not important, because this measurement is
  constrained by the statistics in the appearance channels and
  not by the systematics coming from the matter density uncertainties.
\item[For the $\mbox{\boldmath$\stheta$}$-precision] matter density
  uncertainties are only relevant for $\stheta \gtrsim 10^{-3}$
  because of the statistics domination below that region.
\item[For the sign of $\mbox{\boldmath$\ldm$}$-sensitivity] matter
  density uncertainties are irrelevant because of the very
  distinctive solutions of this binary measurement, which could only
  be mixed up by the matter density uncertainties by using negative
  matter densities.
\item[For the sensitivity to {\em any} CP violation] matter density
  uncertainties again become important for $\stheta \gtrsim
  10^{-3}$. However, their relative influence is reduced for large
  values of $\sdm$.
\item[For the sensitivity to {\em maximal} CP violation] matter
  density uncertainties are not significantly important within
  the KamLAND-allowed range at the $2 \sigma$ confidence level.
\item[For the precision of $\mbox{\boldmath$\deltacp$}$] matter
  density uncertainties are not important for $\stheta \lesssim
  10^{-3}$. However, for $\stheta \gtrsim 10^{-3}$ they make the
  precision worse by a factor of two (close to the CHOOZ limit).
\end{description}

We conclude that three conditions are necessary for the matter density
uncertainties to become relevant at the considered neutrino factory
analysis: First, only some of the measurements are influenced by
matter density uncertainties, such as the $\stheta$- and
$\deltacp$-precision measurements. 
Second, matter density uncertainties are only important for $\stheta
\gtrsim 10^{-3}$.
Third, matter density uncertainties only sensibly affect the measurements for
uncertainty amplitudes $\Delta \rho^* \gtrsim 1 \% \cdot \bar{\rho}$, as
it can be observed in any of the figures so far.

Since superbeams or superbeam upgrades are basically sensitive down
to $\stheta \sim 10^{-3}$ for the discussed measurements, competitiveness
of neutrino factories with superbeams is an interesting issue. 
Superbeams are often proposed with shorter baselines and they are
using different (non-resonant) neutrino energies, which means that they are
much less affected by matter density uncertainties. This is one of the
reasons why they are often much better than the neutrino factories
for $\stheta \gtrsim 10^{-3}$. For illustration, the $3 \sigma$ CP violation 
discovery potential is shown in \figu{cpviol3s} for \NuFactII\ 
(with $\Delta \rho^* = 5 \% \cdot \bar{\rho}$ and $\Delta \rho^* \lesssim
1 \% \cdot \bar{\rho}$) and the JHF to 
Hyper-Kamiokande superbeam upgrade \JHFHK\ from \Ref~\cite{Huber:2002mx}
(as it is proposed in \Ref~\cite{Itow:2001ee}, using two years of
neutrino running and six years of antineutrino running). Since this
superbeam upgrade is proposed with a quite short baseline 
$L \simeq 295 \, \mathrm{km}$, it is hardly affected by matter effects and
matter density uncertainty effects.
The figure clearly demonstrates that especially for $\stheta \gtrsim 10^{-2}$
a knowledge on the matter density profile with about $1\%$ precision
is necessary for the competitiveness of neutrino factories with
superbeams. Therefore, it would be interesting to know with what costs
such a precision could be achieved in geophysics along a specific baseline.
However, if the superbeams do not find $\stheta>0$, then the
neutrino factories will for $\stheta \lesssim 10^{-3}$ be almost
unaffected by matter density uncertainties in the post-superbeam era.

\subsection*{Acknowledgments}

We would like to thank Steve Geer, Patrick Huber, and Manfred Lindner
for useful discussions and comments.
T.O. and W.W. would like to thank TUM and KTH-SCFAB, respectively, for
the warm hospitality during their respective visits at the other university.
This work was supported by the Magnus Bergvall Foundation
(Magn.~Bergvalls Stiftelse), the ESF Network on Neutrino Astrophysics
[T.O.], the Swedish Research Council (Vetenskapsr{\aa}det), Contract
No.~621-2001-1611, 621-2002-3577 [T.O.], the G{\"o}ran Gustafsson
Foundation (G{\"o}ran Gustafssons Stiftelse) [T.O.], the
``Studienstiftung des deutschen Volkes'' (German National Merit
Foundation) [W.W.], and the ``Sonderforschungsbereich 375 f{\"u}r
Astro-Teilchenphysik der Deutschen Forschungsgemeinschaft'' [W.W.].

\end{document}